%% file: main.tex
\documentclass[12pt]{article}
\usepackage{natbib,bm}
\usepackage{amsmath}
\usepackage{amssymb}
\usepackage{graphicx}
\usepackage[dvips]{epsfig}
\usepackage{lscape}
\usepackage{rotating}
\usepackage{mathtools}
\usepackage{setspace}

\renewcommand{\baselinestretch}{1.1}
\tiny\normalsize
\input Macros

\makeatletter
\def\vec{\mathop{\operator@font vec}\nolimits}
\makeatother

\DeclareMathOperator{\diag}{diag}

\begin{document}
\input text

\renewcommand{\baselinestretch}{1.0}

\input{main.bbl}
%\tiny\normalsize
%\bibliographystyle{agsm}
%\bibliography{ref}

\newpage
\input table

\newpage
\input graph

\end{document}

%% file: Macros.tex
\def\c#1{\ensuremath{\mathcal{#1}}}

\newcommand{\nmathbf}{\bm}

\def\bfD{\nmathbf D}

\def\bfV{\nmathbf V}

\def\bfe{\nmathbf e}
\def\bff{\nmathbf f}
\def\bfg{\nmathbf g}

\def\bfq{\nmathbf q}

\def\bft{\nmathbf t}
\def\bfu{\nmathbf u}

\def\bfw{\nmathbf w}
\def\bfx{\nmathbf x}
\def\bfy{\nmathbf y}
\def\bfz{\nmathbf z}

\def\bfbeta   {\nmathbf \beta}
\def\bfgamma  {\nmathbf \gamma}

\def\bftheta  {\nmathbf \theta}

\def\bfmu     {\nmathbf \mu}

\def\bfxi     {\nmathbf \xi}

\def\bfLambda {\nmathbf \Lambda}

\def\bfOmega  {\nmathbf \Omega}

\newcommand{\cfD}{\mbox{\c{D}}}

\newcommand{\cfL}{\mbox{\c{L}}}

\newcommand{\cfT}{\mbox{\c{T}}}

\def\boldfacefake#1{\kern-4pt
   \hbox{ \mathsurround=0pt
   \hbox to 0.4pt{$#1$\hss}\hbox to 0.4pt{$#1$\hss}\hbox {$#1$}}}

\newcommand{\E}{\mbox{E}}

\newcommand{\Cov}{\mbox{Cov}}

\newcommand{\ok}{\hfill\fbox{}}

\newcommand{\btable}{\begin{table}[h]\centering}
\newcommand{\etable}{\end{table}}
\newcommand{\bt}{\begin{parag}\small \let\b=\nsb \let\sb=\nssb \begin{tabular}}
\newcommand{\et}{\end{tabular}\let\b=\nb \let\sb=\nsb\end{parag}}

\newenvironment{parag}{\par}{\par}
\newenvironment{dif}
    {\begin{parag}\small \let\b=\nsb \let\sb=\nssb \begin{parag}}
    {\let\b=\nb \let\sb=\nsb \end{parag}\end{parag}}

\newenvironment{proof}{\begin{dif} \noindent{\em Proof.~}}
            {\ok\vspace*{10pt}\end{dif}}
\newenvironment{exa}{\begin{list}{}
           {\setlength{\leftmargin}{10pt}
            \setlength{\rightmargin}{\leftmargin}}
           \item\begin{ex}\em}{\end{ex}\end{list}}

\newcommand{\be}{\begin{eqnarray}}
\newcommand{\ee}{\end{eqnarray}}
\newcommand{\ba}{\begin{eqnarray*}}
\newcommand{\ea}{\end{eqnarray*}}

\newtheorem{theorem0}{Theorem}
\newtheorem{lemma0}{Lemma}
\newtheorem{remark0}{Remark}
\newtheorem{fact0}{Fact}
\newtheorem{example0}{Example}
\newtheorem{definition0}{Definition}
\newtheorem{corollary0}{Corollary}
\newtheorem{proposition0}{Proposition}
\newtheorem{algorithmY}{Algorithm}

\newcommand{\reals}{\mbox{\rm I\kern-.20em R}}
\newcommand{\sreals}{\mbox{\small \rm I\kern-.20em R}}

\newcommand{\goto}{\rightarrow}

\newcommand{\bdfn}{\begin{dfn}}
\newcommand{\edfn}{\end{dfn}}
\newcommand{\bteo}{\begin{teo}}
\newcommand{\eteo}{\end{teo}}
\newcommand{\bexa}{\begin{exa}}
\newcommand{\eexa}{\end{exa}}
\newcommand{\bdif}{\begin{dif}}
\newcommand{\edif}{\end{dif}}
\newcommand{\bpro}{\begin{proof}}
\newcommand{\epro}{\end{proof}}

%% file: text.tex
\begin{center}
 {\large \bf Bayesian Nonparametric Inference for Panel Count Data with an Informative Observation Process}

\bigskip
 Ye Liang$^1$, Yang Li$^2$ and Bin Zhang$^3$ \\
 {\small $^1$ \it Department of Statistics, Oklahoma State University, Stillwater, OK 74074} \\
 {\small $^2$ \it Department of Mathematics and Statistics, University of North Carolina, Charlotte, NC 28223} \\
 {\small $^3$ \it Division of Biostatistics and Epidemiology, Cincinnati Children's Hospital, Cincinnati, OH 45229}

\end{center}
\medskip

\begin{abstract}
In this paper, the panel count data analysis for recurrent events is considered. 
Such analysis is useful for studying tumor or infection recurrences in both clinical trial and observational studies.
A bivariate Gaussian Cox process model is proposed to jointly model the observation process and the recurrent event process. 
Bayesian nonparametric inference is proposed 
for simultaneously estimating regression parameters, bivariate frailty effects and baseline intensity functions. 
Inference is done through Markov chain Monte Carlo, with fully developed computational techniques. 
Predictive inference is also discussed under the Bayesian setting. The proposed method is shown to be efficient
via simulation studies. A clinical trial dataset on skin cancer patients is analyzed to illustrate the proposed approach.  

\end{abstract}

\section{Introduction} \label{sec:intro}
Panel count data in medical studies often refer to incomplete recurrent event data observed only at finite distinct observation time points. 
The set of observation times may vary from subject to subject. Using mathematical notations, for subject $i$ from a sample of $n$ subjects, 
we observe the subject only at discrete time points: $t_{i,1},\ldots,t_{i,m_i}$, where $m_i$ is the total number of observations for subject $i$. 
At any time point $t_{i,j}$, we observe a cumulative count $N_{i,j}$ of a recurrent event, but the actual event times are unknown. 
A classical example of panel count data is the bladder tumor data \citep{Sun2000}. In the study, a list of post-surgical patients were assigned to
three treatment groups. Each patient had multiple random clinical visits and the number of recurrent tumors between two visits 
were observed. Another example was a chemotherapy trial for skin caner patients \citep{Li2011}. In the study,  
two treatment groups of patients were followed up at clinical visits and the number of recurrent non-melanoma skin cancers 
were observed. Other examples include infection recurrences in leukemia patients and respiratory system exacerbations 
among cystic fibrosis patients \citep{Kal2011}. 

One common approach for modeling panel count data is to regard the underlying recurrent event process as a counting process, 
for instance, the nonhomogeneous Poisson process.
For regression analysis of panel count data, the Cox model is a popular choice \citep{Anderson1982}, 
where covariates and a baseline intensity function are specified in a log-linear form. 
Recently, there has been an increasing attention in literature on the dependence between the observation process and 
the recurrent event process, for which a bivariate joint modeling seems a natural approach.
Such bivariate models usually assume dependence specified through subject-specific frailties.
\cite{He2009, Huang2006, Sun2007} proposed joint modeling approaches that depend on shared random effects.  
\cite{Li2010} proposed a class of semiparametric transformation models. \cite{LiY2015} proposed a semiparametric regression model
in which the underlying dependence structure for random effects are unspecified. 
\cite{Zhao2013b, Zhou2016} combined such context with terminal events. 

For all existing bivariate joint modeling that we are aware of, estimation on regression parameters was primarily focused but 
estimation on baseline intensity functions was rarely considered for the nature of Cox model and its generalizations.
However, a smooth estimate of baseline functions may still be useful in terms of comprehension and prediction, 
to both physicians and patients. \cite{Altman2000, Royston2013} have argued this point for baseline hazard functions in survival analysis. 
On the other hand, previous research work on baseline estimation predominantly used spline-based models 
\citep{Nielsen2008, Lu2007, Lu2009, Hua2012, Hua2014, Yao2016}, but without considering a dependent observation process. 
It remains unclear how these spline-based methods can be extended to cases where the observation process is correlated.  

In this paper, we consider a bivariate joint modeling for panel count data when the observation and event processes are dependent. 
Our main goal is to provide an inferential procedure that can simultaneously estimate regression coefficients and baseline 
intensity functions while allowing for correlated processes. For this purpose, we propose a log-Gaussian Cox process model
under the Bayesian framework, which can be shown more flexible than existing models. In addition, 
we develop a nonparametric Bayesian inference through Markov chain Monte Carlo (MCMC) for the proposed model.  
With the proposed Bayesian inference, we can estimate both the intensity function and the mean function,
and furthermore, do predictive inference on disease recurrences for future patients.

The rest of this paper is organized as follows. Section 2 describes the model specification, especially the log-Gaussian Cox process. 
Section 3 establishes the inference framework, including Bayesian inference and posterior sampling steps.  
Section 4 describes Bayesian predictive inference using posterior samples.  
Section 5 presents results from extensive simulation studies, where our method is compared with several existing methods. 
In Section 6, the proposed method is applied to a clinical trial dataset on a skin cancer treatment.   
Section 7 is a discussion on future directions. 

\section{The Model}
\subsection{The log-Gaussian Cox process model} 
Suppose we observe subject $i$ at distinct time points $t_{ij}$, $j=1,\ldots,m_i$ and $i=1,\ldots,n$. 
At each time point $t_{ij}$, we observe a cumulative count $N_{ij}$ of the recurrent event. We note that the underlying 
observation process $\{T_i(t),t\in \mathbb{R}\}$ and the event process $\{N_i(t),t\in \mathbb{R}\}$ are dependent. 
Suppose $\{T_i(t),t\in \mathbb{R}\}$ follows a nonhomogeneous Poisson process with intensity function $\mu_i(t)$ 
and we consider a Cox regression model,
\be
	\mu_i(t)=\mu_0(t)\exp\{\bfx_i'\bfgamma\}u^O_i, 
\ee
where $\mu_0(t)$ is the baseline intensity function, $\bfx_i$ are covariates and $u^O_i$ is a subject-specific frailty. 
We also assume that the event process $\{N_i(t),t\in \mathbb{R}\}$ is a nonhomogeneous Poisson processes 
with intensity function $\lambda_i(t)$ given by
\be
	\lambda_i(t)=\lambda_0(t)\exp\{ \bfx_i'\bfbeta\}u^N_i,
\ee
where $\lambda_0(t)$ is the baseline intensity function and $u^N_i$ is a frailty that is correlated with $u^O_i$.

Since both $\mu_0(t)$ and $\lambda_0(t)$ are non-negative random functions, it is natural to consider a logarithm transformation
so that the resulting functions can take unrestricted values. Assume that $\mu_0(t)=\exp\{g_1(t)\}$ and $\lambda_0(t)=\exp\{g_2(t)\}$, 
where each of $\{g_1(t): t\in \mathbb{R}\}$ and $\{g_2(t): t\in \mathbb{R}\}$ is a stationary Gaussian process, with constant means $\gamma_0$
and $\beta_0$, covariance functions $C_1(h)=\Cov\{g_1(t),g_1(t+h)\}$ and $C_2(h)=\Cov\{g_2(t),g_2(t+h)\}$, i.e. 
\be
	g_1(t) \sim \mbox{GP}(\gamma_0, C_1(h)) ~~\mbox{and}~~ g_2(t) \sim \mbox{GP}(\beta_0, C_2(h)).
\ee
Notice that $\bfx_i'\bfgamma$ and $\bfx_i'\bfbeta$ should not contain intercept terms to avoid identifiability problems. 
This specification on $\mu_0(t)$ and $\lambda_0(t)$ actually gives us Cox processes, which is defined by
assuming the intensity function of a nonhomogeneous Poisson process from a nonnegative-valued stochastic process.
The Cox process is also known as a doubly stochastic Poisson process. In our proposed model, we in fact have a log-Gaussian 
Cox process \citep{Moller1998}, with the log-intensity function being Gaussian.
Given that spline models have been predominantly used for panel count data analysis,
it should be noted that the equivalence between splines and Gaussian processes is known as early as \cite{Kim1970}. 
Spline models can be viewed as special Gaussian processes with certain kernels. One advantage of Gaussian processes 
is that one no longer needs to consider the knots placement, which, on the other hand, is crucial for spline models
\citep{Rasmussen2006}. 

We give the following moment properties for subject specific intensity functions as follows. Under the specified log-Gaussian
Cox process model, given $\bfgamma$, $\bfbeta$, $u^N_i$ and $u^O_i$, 
\ba
	\E\{\mu_i(t)\} &=& \exp\left\{\bfx_i'\bfgamma+\frac{C_1(0)}{2}\right\}u^O_i, \\
	\E\{\lambda_i(t)\} &=& \exp\left\{\bfx_i'\bfbeta+\frac{C_2(0)}{2}\right\}u^N_i, \\
	\Cov\{\mu_i(t), \mu_i(t+h)\} &=& \left[ \E\{\mu_i(t)\} \right]^2 \left[ \exp\{C_1(h)\}-1 \right], \\
	\Cov\{\lambda_i(t), \lambda_i(t+h)\} &=& \left[ \E\{\lambda_i(t)\} \right]^2 \left[ \exp\{C_2(h)\}-1 \right].
\ea

\subsection{Frailty model}
A correlated frailty model is commonly used to model the bivariate dependence \citep{Sun2007, Li2011, Zhao2013a, LiY2015}. 
A link function between $u^O_i$ and $u^N_i$ is often assumed, for example, $u^N_i=(u^O_i)^{\alpha}$, 
where $\alpha$ apparently controls the dependence. One limitation is that the link function needs to be pre-specified
and how to choose a good one is unclear. Another arguably more serious limitation, is that 
a link function may lead to unappealing properties. For the link function considered above, 
the parameter $\alpha$ will be zero in the case of independence, which forces $u^N_i=1$ for all subjects
and thus fails to model the subject-dependent variation. 

A better alternative is to consider a bivariate distribution for $(u^O_i, u^N_i)$. Let $(\log u^O_i, \log u^N_i)'=(z_{1i},z_{2i})'=\bfz_i$ be 
a bivariate normal distribution $\bfz_i \sim N_2({\bf0}, \bfD)$, where the mean is restricted to be zero to avoid identifiability problems. 
Equivalently, $(u^O_i, u^N_i)'$ follows a bivariate lognormal distribution \citep{Mostafa1964}. 
Following a straightforward calculation, the cross covariance between $\mu_i(t)$ and $\lambda_i(t)$, given $\bfgamma$ and $\bfbeta$
is now
\ba
	\Cov\{\mu_i(t), \lambda_i(t+h)\} = \E\{\mu_i(t)\} \E\{\lambda_i(t)\} (e^{D_{12}} -1),
\ea
in which, $D_{12}$ controls the dependence between the two processes. 

\subsection{Covariance functions and hyper-priors}
The theoretical requirement for the covariance function $C(h)$ is that any covariance
matrix constructed by it be positive-definite. Let $C(h)=\sigma^2 r(h)$, 
where $r(h)$ is a parametric correlation function (or called a Gaussian kernel).  
The Gaussian process is flexible with various choices of kernels. We consider the widely used Mat\'ern kernel: 
\ba
	r(h)=1/2^{\nu-1}\Gamma(\nu) \left(| h |/\theta \right)^{\nu}K_{\nu} \left(| h |/\theta \right),
\ea
where $K_{\nu}(\cdot)$ is the modified Bessel function of order $\nu$ and $\theta$ is the scale parameter. 
The shape parameter $\nu$ is often pre-specified for a desired differentiability of the curve. When $\nu=1/2$, the Mat\'ern
kernel reduces to the exponential kernel $r(h)=\exp\{-|h|/\theta\}$, and when $\nu \goto \infty$, 
the Mat\'ern kernel goes to the squared exponential kernel $r(h)=\exp\{-|h|^2/\theta^2\}$. 

Lastly, we specify hyper-priors for the remaining parameters in the model. Let the regression coefficients have noninformative priors: 
$\pi(\bfgamma) \propto 1$ and $\pi(\bfbeta) \propto 1$. Let the frailty covariance $\bfD$ be inverse-Wishart: $\bfD \sim \mbox{IWish}(k_0, \bfV_0)$, 
with $k_0=3$ and $\bfV_0=\diag\{1\}$, so that this prior is only weakly informative. For hyperparameters in $C_1(h)$ and $C_2(h)$, 
let $\sigma_k^2$, $k=1,2$, be inverse gamma: $\sigma_k^2 \sim \mbox{IG}(a_0, b_0)$, with $a_0=b_0=1$. 
The priors on the length-scales $\theta_k$ are chosen to be informative gamma priors based on empirical evidences \citep{Diggle2013}. 

\section{Bayesian Inference}
\subsection{The joint posterior distribution}
With the full Bayesian model specified in Section 2, we now give inference and computation details. 
For each subject $i$, we observe a complete realization of the observation process $T_i(t)$: $t_{i,1},\ldots,t_{i,m_i}$, 
up to a given follow-up time $C_i$. Under the Poisson process assumption, the likelihood function for observation times is
\ba
	L_1(\mu_0(\cdot), \bfgamma, \bfw; \bft ) &=& \prod_{i=1}^n m_i! \, p(m_i)p(t_{i,1},\ldots,t_{i,m_i} \mid m_i) \\
	&=& \prod_{i=1}^n \left\{ e^{-\int_{0}^{C_i} \mu_0(s)\exp\{\bfx_i'\bfgamma\}u^O_i ds} \prod_{j=1}^{m_i} \mu_0(t_{ij})\exp\{\bfx_i'\bfgamma\}u^O_i \right\}.
\ea
Then for each event process $N_i(t)$, we observe cumulative counts given a realized $T_i(t)$. Let $y_{ij}=N_{ij}-N_{i,j-1}$ be the increment 
at time $t_{ij}$. The likelihood function for the increments, conditional on $\bft$, is
\ba
	L_2( \lambda_0(\cdot),\bfbeta,\bfu; \bfy \mid \bft ) &=& \prod_{i=1}^n\prod_{j=1}^{m_i} \mbox{Poi}\left(y_{ij}; \int_{t_{i,j-1}}^{t_{ij}}
		\lambda_0(s)\exp\{ \bfx_i'\bfbeta\}u^N_i ds \right) \\
	&=& \prod_{i=1}^n \prod_{j=1}^{m_i} \left[ \left\{ \int_{t_{i,j-1}}^{t_{ij}} \lambda_0(s)\exp\{ \bfx_i'\bfbeta\}u^N_i ds \right\}^{y_{ij}} 
		e^{-\int_{t_{i,j-1}}^{t_{ij}}\lambda_0(s)\exp\{ \bfx_i'\bfbeta\}u^N_i ds} / y_{ij}! \right] \,.
\ea
Then the joint likelihood function $L( \mu_0(\cdot), \lambda_0(\cdot), \bfgamma, \bfbeta, \bfu^O, \bfu^N; \bft, \bfy ) = L_1 \times L_2$. 
With priors specified in Section 2, the joint posterior distribution is given by
\ba
	&& p(\bfgamma, \bfbeta, \bfu^O, \bfu^N, \bfD, g_1(\cdot), g_2(\cdot), \sigma_1^2, \sigma_2^2, \theta_1, \theta_2 \mid \bft, \bfy) \propto 
	L \times \pi(\bfgamma) \times \pi(\bfbeta) \times \pi(\bfu^O, \bfu^N \mid \bfD) \\
	&& ~~~~~~
	\times \pi(\bfD) \times \mbox{GP}(g_1(\cdot) \mid \sigma_1^2, \theta_1) \times \mbox{GP}(g_2(\cdot) \mid \sigma_2^2, \theta_2) \times 
	\pi(\sigma_1^2, \sigma_2^2, \theta_1, \theta_2).
\ea

\subsection{Gibbs sampling}
Inference on the posterior distribution is done computationally using MCMC. Consider a Gibbs sampling
for the overall procedure. The full conditional distributions are discussed as follows.

For regression coefficients $\bfgamma$ and $\bfbeta$, the full conditional distributions are
\be \label{gammafullcond}
	p(\bfgamma \mid \mbox{rest}) \propto \exp\left\{ \sum_{i=1}^n \left( m_i \bfx_i'\bfgamma - e^{\bfx_i'\bfgamma}u^O_i 
		\int_0^{C_i} e^{g_1(s)} ds \right) \right\}
\ee
and 
\be \label{betafullcond}
	p(\bfbeta \mid \mbox{rest}) \propto \exp\left\{ \sum_{i=1}^n \left( N_{i,m_i}\bfx_i'\bfbeta
		- e^{\bfx_i'\bfbeta}u^N_i \int_0^{t_{i,m_i}} e^{g_2(s)} ds \right) \right\}.
\ee
It is straightforward to show that both (\ref{gammafullcond}) and (\ref{betafullcond}) are log-concave. 
Hence the adaptive rejection sampling (ARS) \citep{Gilks1992} applies. 

The full conditional distributions for the intercepts $\gamma_0$ and $\beta_0$ are
\be
	\gamma_0 \mid \mbox{rest} \sim \mbox{N} \left(\frac{{\bf1}'\bfOmega_{1}\bfg_1}{{\bf1}'\bfOmega_{1}{\bf1}}, \frac{1}{{\bf1}'\bfOmega_{1}{\bf1}} \right)
	~~\mbox{and}~~
	\beta_0 \mid \mbox{rest} \sim \mbox{N} \left(\frac{{\bf1}'\bfOmega_{2}\bfg_1}{{\bf1}'\bfOmega_{2}{\bf1}}, \frac{1}{{\bf1}'\bfOmega_{2}{\bf1}} \right)
\ee
where ${\bf1}$ is the vector of ones and $\bfOmega_{k}$, $k=1,2$, is the corresponding covariance matrix for $g_k(\cdot)$, which 
will be discussed in the next sub-section. 

For the frailties $u^O_i$ and $u^N_i$ and their covariance matrix $\bfD$, we instead update $\bfz_i$ to avoid the lognormal density. 
Then the full conditional distribution for $\bfz_i$ is
\be
	 p(\bfz_i \mid \mbox{rest}) \propto \exp\left\{ -\frac{1}{2} \bfz_i' \bfD^{-1} \bfz_i 
	+ z_{1i}m_i + z_{2i}N_{i,m_i} - e^{z_{1i}+\bfx_i'\bfgamma}\int_0^{C_i} e^{g_1(s)} ds -e^{z_{2i}
	+\bfx_i'\bfbeta}\int_0^{t_{i,m_i}} e^{g_2(s)} ds \right\}. 
\ee
Again, it is straightforward to show the log-concavity of this conditional density and the ARS is used for sampling $\bfz_i$. 
The full conditional distribution for $\bfD$ is
\be
	\bfD \mid \mbox{rest} \sim \mbox{Inv-Wishart} \left(n+k_0, \sum_{i=1}^n \bfz_i \bfz_i' +\bfV_0 \right).
\ee

For sampling the latent Gaussian process components $g_1(\cdot)$ and $g_2(\cdot)$, and their hyperparameters $\sigma_1^2$, 
$\sigma_2^2$, $\theta_1$ and $\theta_2$, we have the following full conditional distributions:
\be
	p(g_1(\cdot) \mid \mbox{rest}) \propto \mbox{GP}(g_1(\cdot) \mid \sigma_1^2, \theta_1) 
		\left\{ \prod_{i=1}^{n} \prod_{j=1}^{m_i} e^{g_1(t_{ij})} \right\} 
		\exp\left\{ -\sum_{i=1}^n e^{\bfx_i'\bfgamma}u^O_i \int_0^{C_i} e^{g_1(s)} ds \right\}, \label{gfullcond1}	
\ee
and
\be
	p(g_2(\cdot) \mid \mbox{rest}) \propto \mbox{GP}(g_2(\cdot) \mid \sigma_2^2, \theta_2) 
		\left\{ \prod_{i=1}^{n} \prod_{j=1}^{m_i} \left( \int_{t_{i,j-1}}^{t_{ij}} e^{g_2(s)}ds \right)^{y_{ij}} \right\} 
		\exp\left\{ -\sum_{i=1}^n e^{\bfx_i'\bfbeta}u^N_i \int_0^{t_{i,m_i}} e^{g_2(s)} ds \right\}. \label{gfullcond2}
\ee
This requires high dimensional sampling and will be discussed in the next sub-section.  

\subsection{Riemann manifold Hamiltonian Monte Carlo}
Since $g_1(\cdot)$ and $g_2(\cdot)$ are infinitely dimensional, a finely spaced grid over the time region of interest (i.e. a discretization) 
is needed. For instance, consider a grid $\{s_l: l=0, \ldots, L\}$ that covers the entire time period. A pre-defined grid will induce a finite 
multivariate normal distribution from the Gaussian process, say, $\bfg_1=(g_1(1),\ldots,g_1(L))' \sim \mbox{N}_L(\bfmu_{g_1}, \bfOmega_1)$ 
and $\bfg_2=(g_2(1),\ldots,g_2(L))' \sim \mbox{N}_L(\bfmu_{g_2}, \bfOmega_2)$. 
Without loss of generality, assume that the entire time region has length one. Then the cell length of the grid is $1/L$ 
and an observation time $t_{ij}$ is approximated as its corresponding cell index: $\lceil t_{ij}L \rceil$.
The intensity function within each cell is approximated by a constant function, and then any integral in (\ref{gfullcond1}) and (\ref{gfullcond2}) 
is approximated by
\ba
	\int_{t_{i,j-1}}^{t_{ij}} e^{g_2(s)}ds = \frac{1}{L}\sum_{l=\lceil t_{i,j-1}L \rceil}^{\lceil t_{ij}L \rceil} e^{g_2(l)}. 
\ea
Note that the choice of grid and $L$ is completely arbitrary. The fineness of such grid only reflects a balance between computational 
complexity and the accuracy of the approximation \citep{Diggle2013}. 

Consider the logarithm of the approximated conditional density of either (\ref{gfullcond1}) or (\ref{gfullcond2}), denoted by $\cfL(\bfg)$. 
For a chosen dimension $L$, now it becomes a challenging problem to sample the high dimensional $\bfg$ 
from an irregular $\cfL(\bfg)$. While the traditional Metroplis-Hastings algorithms have low efficiency in such problems, 
there has been an increasing focus on gradient-based sampling techniques, such as Metroplish adjusted Langevin
algorithm and Hamiltonian Monte Carlo \citep{Neal2010}. Despite efficiency gained in those algorithms, they both require careful
tuning in the implementation. A recent promising sampling technique called Riemann manifold Hamiltonian Monte Carlo (RMHMC) 
\citep{Girolami2011} greatly eases the burden of tuning. The RMHMC utilizes the Riemann geometry of the parameter space 
by incorporating a metric tensor, usually the Fisher information matrix, to the Hamiltonian dynamics. 
To show the efficiency of utilizing RMHMC in our application, we compare it with another widely used algorithm for Gaussian process
sampling, called elliptical slice sampling. Figure \ref{fig:rmhmc} shows that the RMHMC converges almost immediately.   

To implement the algorithm, analytical forms of the gradient $\nabla_{\bfg}\cfL(\bfg)$ and the Fisher information 
matrix $-\E_{\bfy,\bfg \mid \bftheta}[\nabla^2_{\bfg}\cfL(\bfg)]$ are both required. 
In our case, the gradient is analytically available for the approximated log-densities $\cfL_1(\bfg_1)$ and $\cfL_2(\bfg_2)$:
\ba
	\nabla_{\bfg_1}\cfL_1(\bfg_1) = -\bfOmega_1(\bfg_1 - \bfmu_{g_1}) + \bff_1 + \bfq_1\circ\bfe_1,
\ea
where $\circ$ is the Hadamard product, $\bfe_1=L^{-1}(e^{g_1(1)}, \ldots, e^{g_1(L)})'$, both $\bff_1$ and $\bfq_1$ are $L$-dimensional vectors with
\ba
	f_{1,l} &=& \sum_{i=1}^n \sum_{j=1}^{m_i}I(l=\lceil t_{ij}L \rceil), \\
	q_{1,l} &=& -\sum_{i=1}^n \exp(\bfx_i'\bfgamma)u^O_i I(l \leq \lceil C_i L \rceil),
\ea
for $l=1,\ldots,L$. On the other hand, we have
\ba
	\nabla_{\bfg_2}\cfL_2(\bfg_2) = -\bfOmega_2(\bfg_2 - \bfmu_{g_2}) + {\bf0} + \bfq_2\circ\bfe_2,
\ea
where $\bfe_2=L^{-1}(e^{g_2(1)}, \ldots, e^{g_2(L)})'$, and $\bfq_2$ is an $L$-dimensional vector with
\ba
	q_{2,l} = L\sum_{i=1}^n\sum_{j=1}^{m_i} \frac{y_{ij}I(\lceil t_{i,j-1}L \rceil \leq l \leq \lceil t_{ij}L \rceil)}{\sum_{l=\lceil t_{i,j-1}L \rceil}^{\lceil t_{ij}L \rceil} e^{g_2(l)}} 
	-\sum_{i=1}^n \exp(\bfx_i'\bfbeta)u^N_i I(l \leq \lceil t_{i,m_i}L \rceil),
\ea
for $l=1,\ldots,L$. For either $\cfL_1$ or $\cfL_2$, the Fisher information matrix is analytically too complicated, but can eventually
be expressed as $-\E_{\bfy,\bfg \mid \bftheta}[\nabla^2_{\bfg}\cfL(\bfg)]=\bfOmega^{-1}+\bfLambda/L$,
where $\bfLambda$ has a complicated form. As $L$ goes large, the second term is negligible. 
Therefore, $\bfOmega^{-1}$ roughly is the desired metric tensor and is used as the mass matrix 
in Hamiltonian dynamics for both cases. Note, even if $\bfOmega^{-1}$ is not sufficiently close to the actual metric tensor, the HMC algorithm
is still valid because the mass matrix can be any matrix in theory. The rough use is only a matter of efficiency, not validity.  

Hyperparameters $\sigma_1^2$, $\sigma_2^2$, $\theta_1$ and $\theta_2$ in the Gaussian process are sampled alternately 
from $p(\sigma_k^2 \mid \mbox{rest}) \propto N(\bfg_k \mid \sigma_k^2, \theta_k)\pi(\sigma_k^2)$ and 
$p(\theta_k \mid \mbox{rest}) \propto N(\bfg_k \mid \sigma_k^2, \theta_k)\pi(\theta_k)$, $k=1,2$, where 
the former one is inverse gamma and the latter one is sampled by the adaptive rejection Metroplis sampling. 

\section{Prediction}
Prediction is often considered an important part in ordinary linear regressions and generalized linear regressions. 
It is rarely discussed in regression analysis for panel count data. One advantage of Bayesian analysis is that the
predictive inference automatically accounts for parameter uncertainties. Consider a future subject with covariates $\tilde{\bfx}$. 
We are interested in a predictive distribution of the count $\tilde{y}$ during a pre-specified time period $\cfT$.
Let $\cfD$ represent all past data and let $\bfxi$ represent all parameters in the model. Then the posterior predictive 
distribution for $\tilde{y}$ is
\ba
	p(\tilde{y} \mid \cfD, \tilde{\bfx})=\int_{\cfD} p(\tilde{y} \mid \bfxi, \tilde{\bfx}) p(\bfxi \mid \cfD) d\bfxi.
\ea
As long as we have posterior samples from $p(\bfxi \mid \cfD)$, this predictive distribution is computationally available. 
Suppose we have $b=1,\ldots, B$ posterior samples from MCMC, that is, we have $\bfxi^{(b)}$ containing 
$\lambda_0^{(b)}(\cdot), \bfbeta^{(b)}$ and $\bfD^{(b)}$. We can proceed as follows: for each $b$, 
\begin{enumerate}
	\item Draw $\tilde{u}^{(b)}$ from $\mbox{lognormal}(0, D_{22}^{(b)})$.  
	\item Draw $\tilde{y}^{(b)}$ from $\mbox{Poi}(\tilde{E}^{(b)})$, with 
	$\tilde{E}^{(b)} = \exp\{\tilde{\bfx}' \bfbeta^{(b)}\}\tilde{u}^{(b)} {\displaystyle \int_{\small \cfT} \lambda_0^{(b)}(t) dt}$.  
\end{enumerate}  
Then, $\tilde{y}^{(1)}, \ldots, \tilde{y}^{(B)}$ is a sample from the correct predictive distribution. Note that the prediction is made
marginally on the event process as the observation time period $\cfT$ is pre-specified. Prediction of observation times 
is rarely of interest, but possible under this model. A medical practitioner may be interested in a probability of 
no recurrence in $\cfT$, given the subject's covariates, and that is $P(\tilde{y}=0 \mid \cfD, \tilde{\bfx})$, computed from the predictive distribution.
  
\section{Simulation}
Our proposed Bayesian inference can simultaneously estimate regression coefficients and baseline intensity functions with allowing 
for correlated processes. There was no previous work that handles such a complicated scenario in a single model. In this simulation section,
we compare our proposed inference with existing approaches which give different types of partial results.
The simulation section is designed as two sub-themes. Theme 1 is that we compare our method with 
HSW \citep{Hu2003} and ZTS \citep{Zhao2013a} on estimation of regression parameters. 
Both work focused on estimating regression parameters alone without estimating baseline functions. 
Theme 2 is that we compare our method with YWH \citep{Yao2016} on estimation of cumulative baseline functions. 
YWH used monotone splines for cumulative baseline functions, however, did not consider a dependent observation process. 

\subsection{Comparing estimation of regression coefficients}
In Theme 1, we compare estimation of regression parameters between our method and those from HSW and ZTS 
under two simulation settings: 
\begin{enumerate}
	\item [1.] Setting 1 represents an original case that was considered in ZTS, where our model and the one in HSW 
		are both misspecified. Baseline intensity functions are 
		$\mu_0(t)=1/8$ and $\lambda_0(t)=1/t$. The censoring times $C_i$ are generated from $\mbox{Unif}(2,9)$.
		The frailties are generated as $u_i^O \sim \mbox{Ga}(2, 0.2)$ and $u_i^N=(u_i^O)^{0.5}+\mbox{Ga}(1,2)$. 
		Let $x_i$ be from Bernoulli with probability $0.5$ and let $\beta=\gamma=1$. Sample size is set to be $n=100$. 
	\item [2.] Setting 2 represents a case for which none of the three models is misspecified. 
		Baseline intensity functions are $\mu_0(t)=0.25 \{ \exp(-t/20) + 0.5\exp[-((t-70)/40)^2]\}$
		and $\lambda_0(t)=0.125 \{ \exp(-t/10) + 0.5\exp[-((t-70)/20)^2]\}$. 
		The censoring times $C_i$ are generated from $\mbox{Unif}(50,100)$.
		The frailties are generated independently from lognormal with zero mean and $0.25$ variance.  
		Let $x_i$ be from $\mbox{Unif}(0,1)$ and let $\beta=\gamma=1$. Sample size is set to be $n=100$. 
\end{enumerate}
For each setting we generated $500$ datasets and applied all three methods to the same datasets. 
For our proposed method, we set $\nu=2.5$ in the Mat\'ern kernel as a balanced choice for differentiability. 
To obtain fast convergence in each replication of the simulation study, we pre-fixed $\theta_1=\theta_2=0.5$ for Setting 1 
and $\theta_1=\theta_2=4$ for Setting 2. The choice of hyperparameters will not drastically change the estimation, 
as shown in a sensitivity analysis in Section 6. 
In each replication of the $500$ datasets, we ran the MCMC algorithm for $20,000$ iterations with a burn-in size of $5,000$.
Since $\beta$, of the event process, is usually of primary interest, and note that $\gamma$ is not available in ZTS, 
we compare the estimated bias, root mean squared errors (RMSE) and 95\% coverage probabilities (CP) for $\beta$ only. 
The comparison results are presented in Table \ref{tab:sim1}. It can be seen that all three methods result in small biases, 
reasonable RMSE and CP. However, our proposed method outperforms HSW and ZTS under both settings. 

\subsection{Comparing estimation of baseline functions}
In Theme 2, we compare our method with the method in YWH, in particular on estimation of cumulative baseline functions.
Our method provides both intensity and cumulative intensity estimation under dependent frailties, and hence is
more general than YWH. We used the R package \textsf{``PCDSpline''} developed by YWH to implement their method. 
Consider the following setting for simulating panel count data. 
\begin{enumerate}
	\item [3.] Baseline intensity functions are $\mu_0(t)=0.25 \exp\{-t/100\}$ and 
		$\lambda_0(t)=0.25\exp\{-(t-20)^2/25\} + 0.25\exp\{-(t-50)^2/25\} + 0.25\exp\{-(t-80)^2/25\}$. 
		The censoring times $C_i=100$ are fixed for all subjects. 
		The frailties are generated from a bivariate lognormal with zero mean and covariance matrix $D_{11}=D_{22}=0.25$ and $D_{12}=0.125$. 
		Let $x_{1i}$ and $x_{2i}$ both be from $\mbox{Unif}(0,1)$ and let $\beta_1=\gamma_1=-1$ and $\beta_2=\gamma_2=1$. 
		Sample size is set to be $n=100$.
\end{enumerate}
We generated $500$ datasets and applied both methods to the same datasets.
For the method in YWH, we used both linear bases and quadratic bases for the spline model, with 10 equally spaced knots 
following the recommendation in YWH. For our proposed method, we set $\nu=2.5$ and pre-fixed $\theta_1=4$ and $\theta_2=2$ 
for the same reasons stated in Section 5.1. In each replication of the $500$ datasets, 
we ran the MCMC algorithm for $20,000$ iterations with a burn-in size of $5,000$. We compare estimates on regression parameter 
$\bfbeta$ and a rescaled cumulative intensity function $\Lambda_0(t)=\int_0^t \lambda_0(s)ds / \int_0^{\cfT} \lambda_0(s)ds$ 
over the time region $[0, \cfT]$. Table \ref{tab:sim2} shows biases, RMSE and CP for $\bfbeta$ and $\Lambda_0(t)$ on four interior time points 
within the region $[0, 100]$. Figure \ref{fig:sim1} shows point-wise comparisons of biases and RMSE over the region $[0, 100]$. 
It can be seen that our proposed method performs better with smaller biases and RMSE, and more reasonable CP than those given by YWH. 
In addition, Figure \ref{fig:sim2} shows averaged estimates of $\lambda_0(t)$, $\mu_0(t)$ and their cumulative functions 
based on 500 replicates, together with their true curves. These averaged estimates match their true curves extremely well. 

\section{Skin Cancer Data}
The skin cancer data were analyzed in \cite{Li2011} and \cite{Yao2016}. A chemoprevention trial for skin cancer patients were conducted 
by the University of Wisconsin Comprehensive Cancer Center. The aim was to evaluate the effectiveness of the 0.5 g/m$^2$/day DFMO 
treatment in reducing recurrent tumors for patients with non-melanoma skin cancers.
The study consists of 291 patients (with one removal) in total, who were then randomized into a placebo group (147) and a treatment group (144). 
Two types of cancer, basal cell carcinoma and squamous cell carcinoma, were combined in this analysis. 
Two covariates, the treatment indicator and the number of initial tumors, were included in this analysis. 
We chose hyperparameters in the GP as follows: let $\nu=1.5$ in the Mat\'ern kernel and $\theta_k \sim \mbox{Ga}(4,4)$ for $k=1,2$. 
We ran MCMC with randomly picked initial values for $200,000$ iterations with a $50,000$ burn-in size.
Convergence diagnostics are given in a supplementary file.  

Regression parameters are estimated as follows: posterior mean (s.d.) of $\beta_1$ (DFMO) is $-0.104~(0.149)$ 
and posterior mean (s.d.) of $\beta_2$ (Initial tumor number) is $0.111~(0.012)$, which indicates that the DFMO treatment
effect is marginal but the initial number of tumors has a significant positive effect on tumor recurrences.  
These numbers are in general consistent with conclusions in \cite{Li2011}. 
Figure \ref{fig:skin1} shows posterior means and credible bands of $\lambda_0(t)$ and $\mu_0(t)$. 
Comparing the estimated observation intensity with the sample observation times, our estimate well represents the shape and pattern in sample data. Notice that there is a clear periodic pattern in observation times, which is due to that the original clinical trial planned scheduling 
follow-up times every six months, however, each patient's actual visit times appear to be random and do not match the schedule exactly. 

Though the same data have been analyzed in literature, results for prediction do not seem available.
Suppose that we are interested in predicting the recurrence count within five years after the initial treatment 
for a future subject, and we are also interested in the probability of no recurrence, a.k.a. free of disease, 
within five years after treatment. We argue that this is an important piece of information for both physicians and patients.
Figure \ref{fig:skin2} shows Bayesian predictive distributions for various combinations of covariates. 
Figure \ref{fig:skin3} shows Bayesian predictive probabilities of the five-year-disease-free event. 

As a model checking step, we considered a sensitivity analysis for difference choices of hyperparameters. 
We considered three different settings: (1) $\nu=2.5$, $\theta_k\sim\mbox{Ga}(8,4)$; 
(2) $\nu=1.5$, $\theta_k\sim\mbox{Ga}(4,4)$; and (3) $\nu=0.5$, $\theta_k\sim\mbox{Ga}(16,4)$, $k=1,2$.
For each setting, we ran MCMC for $200,000$ iterations with a $50,000$ burn-in size. 
Estimates of $\bfbeta$, $\bfgamma$ and $\Lambda_0(t)$ are compared in Table \ref{tab:skin} and Figure \ref{fig:skin4}. 
The results show that the estimation is quite robust with respect to the choice of hyperparameters. 
These choices, however, do change the differentiability and smoothness of the estimated curves. 
To choose between different hyperparameter values, one may consider model selection criteria, such as
the deviance information criterion (DIC). We report the DIC in Table \ref{tab:skin} and in this comparison,
Choice 2 is preferred with the smallest DIC value. 

\section{Discussion}
In this paper, we developed a bivariate log-Gaussian Cox process model for panel count data. We derived inference and computation
procedures for the proposed model. We emphasized the need of smooth estimation of intensity functions and discussed prediction 
in panel count data analysis. One issue of using the Gaussian process model is learning the scale parameter $\theta$ in
the kernel, which determines smoothness of the Gaussian process. It is known that the point process data often provide only weak 
information about $\theta$. We used a fully Bayesian solution with informative priors so that the posterior is balanced between
data information and user's information. On the other hand, an empirical Bayes approach, by maximizing the marginal likelihood of $\theta$, 
is worth future investigations.  

A potential advantage of using Gaussian process models for the baseline intensity function is that
this specification can be useful for modelling multivariate panel count data. Consider that a patient is monitored
for multiple events simultaneously, which is not uncommon in a medical study. For instance, in a clinical trial for assessing influenza vaccines \citep{Zaman2008},
each patient was monitored for multiple symptoms, such as fever, cough and diarrhea. Since all these symptoms are influenza-related,
the underlying intensity patterns may be highly correlated. If we assume $K$ nonhomogeneous Poisson processes for the multiple events,
each intensity function can be modelled by $\lambda_{ik}(t)=\lambda_{0k}(t)\exp\{\bfx_i' \bfbeta_k\}u_{ik}, k=1,\ldots,K$ and $i=1,\ldots,n$. 
To model the underlying dependence between multiple events, let $g_k(t)=\log \lambda_{0k}(t)$ and consider that $(g_1(t),\ldots,g_K(t))$
jointly be a multivariate Gaussian process, with the cross-covariance function between $g_j(t)$ and $g_k(t)$ being
$C_{jk}(h)=\Cov\{g_j(t), g_k(t+h)\}$. Then, the cross-covariance between $\lambda_{ij}(t)$ and $\lambda_{ik}(t)$ (for subject $i$) becomes
\ba
	\mbox{Cov}\{\lambda_{ij}(t), \lambda_{ik}(t+h)\} = \mbox{E}\{\lambda_{ij}(t)\} \mbox{E}\{\lambda_{ik}(t)\} \left[ \exp\{C_{jk}(h) + D_{jk}\} -1 \right],
\ea
where $C_{jk}(h)$ controls the degree of similarity between the two intensity functions and $D_{jk}$ controls the dependence of the overall intensity magnitude. 
In the multivariate extension described above, the Gaussian process specification is useful to accommodate various dependence structures. 

\section*{Acknowledgement}
The computing for this project was performed at the OSU High Performance Computing Center at Oklahoma State University supported in part through the National Science Foundation grant OCI-1126330. We thank an associate editor and a referee for their valuable comments to the earlier version of this manuscript.

%% file: table.tex
\begin{table}[h]
\caption{Simulation: Compare regression parameter estimation with HSW(2003) and ZTS(2013).
	RMSE: root mean squared error. CP: 95\% coverage probability. Bayesian approach uses a symmetric credible interval.}
\label{tab:sim1}
\begin{center}
\begin{tabular}{c|c|c|c|c|c|c|c|c|c|c|c}
\hline
 & \multicolumn{9}{c|}{Setting 1} \\
\hline
 & \multicolumn{3}{c|}{Proposed} & \multicolumn{3}{c|}{HSW} & \multicolumn{3}{c|}{ZTS}  \\
\hline
 & Bias & RMSE & CP & Bias & RMSE & CP & Bias & RMSE & CP \\
\hline
$\beta=1$ (Event) & -0.004 & 0.097 & 0.94 & 0.005 & 0.195 & 0.96 & 0.006 & 0.221 & 0.92  \\
\hline
\hline
 & \multicolumn{9}{c|}{Setting 2} \\
\hline
 & \multicolumn{3}{c|}{Proposed} & \multicolumn{3}{c|}{HSW} & \multicolumn{3}{c|}{ZTS}  \\
\hline
 & Bias & RMSE & CP & Bias & RMSE & CP & Bias & RMSE & CP \\
\hline
$\beta=1$ (Event) & -0.002 & 0.255 & 0.95 & 0.017 & 0.328 & 0.96 & 0.018 & 0.332 & 0.96  \\
\hline
\end{tabular}
\end{center}
\end{table}%

\begin{table}[h]
\caption{Simulation Setting 3: Compare with YWH(2016). 
	RMSE: root mean squared error. CP: 95\% coverage probability. Bayesian approach uses a symmetric credible interval.}
\label{tab:sim2}
\begin{center}
\begin{tabular}{c|c|c|c|c|c|c|c|c|c|c|c}
\hline
 & \multicolumn{3}{c|}{Proposed} & \multicolumn{3}{c|}{YWH Linear} & \multicolumn{3}{c|}{YWH Quadratic}  \\
\hline
 & Bias & RMSE & CP & Bias & RMSE & CP & Bias & RMSE & CP \\
\hline
$\beta_1=-1$ & 0.012 & 0.226 & 0.96 & 0.015 & 0.236 & 0.85 & 0.007 & 0.238 & 0.85 \\
$\beta_2=1$ & 0.002 & 0.238 & 0.95 & -0.015 & 0.235 & 0.76 & -0.005 & 0.238 & 0.76 \\
$\Lambda_0(20)$ & 0.0004 & 0.016 & 0.94 & 0.036 & 0.039 & - & -0.056 & 0.056 & - \\
$\Lambda_0(40)$ & 0.0009 & 0.017 & 0.97 & -0.000 & 0.017 & -  & 0.027 & 0.032 & - \\
$\Lambda_0(60)$ & -0.004 & 0.018 & 0.94 & -0.016 & 0.024 & -  & -0.032 & 0.037 & - \\
$\Lambda_0(80)$ & -0.002 & 0.019 & 0.95 & -0.051 & 0.054 & - & 0.047 & 0.048 & - \\
\hline
\end{tabular}
\end{center}
\end{table}%

\begin{table}[h]
\caption{Skin cancer data: Sensitivity analysis and DIC. Compare hyperparameter choices: 
	(1) $\nu=2.5$, $\theta_k\sim\mbox{Ga}(8,4)$; (2) $\nu=1.5$, $\theta_k\sim\mbox{Ga}(4,4)$; and (3) $\nu=0.5$, $\theta_k\sim\mbox{Ga}(16,4)$, 
	$k=1,2$. Table shows posterior means and standard deviations.}
\label{tab:skin}
\begin{center}
\begin{tabular}{c|c|c|c|c|c}
\hline
 & $\beta_1$ & $\beta_2$ & $\gamma_1$ & $\gamma_2$ & DIC \\
\hline
Choice 1 & -0.105 (0.149) & 0.111 (0.012) & -0.0374 (0.0462) & 0.00844 (0.00407) & 20783.1 \\
Choice 2 & -0.104 (0.149) & 0.111 (0.012) & -0.0370 (0.0462) & 0.00846 (0.00407) & 20766.0 \\
Choice 3 & -0.102 (0.149) & 0.111 (0.012) & -0.0373 (0.0462) & 0.00848 (0.00409) & 20799.2 \\
\hline
\end{tabular}
\end{center}
\end{table}%

%% file: graph.tex
\begin{figure}[ht]
\begin{center}
\includegraphics[scale=0.65,angle=270]{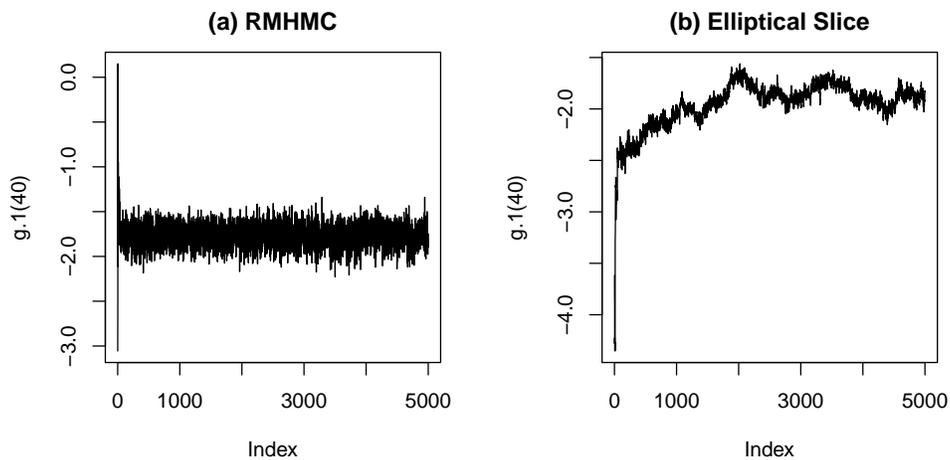}
\caption{(a). Convergence of $g_1(40)$ using RMHMC. (b). Convergence of $g_1(40)$ using elliptical slice sampling. 
	Two sampling techniques are compared with random initial values for the first $5000$ iterations using the skin cancer data.}
\label{fig:rmhmc}
\end{center}
\end{figure}

\begin{figure}[ht]
\begin{center}
\includegraphics[scale=0.7,angle=270]{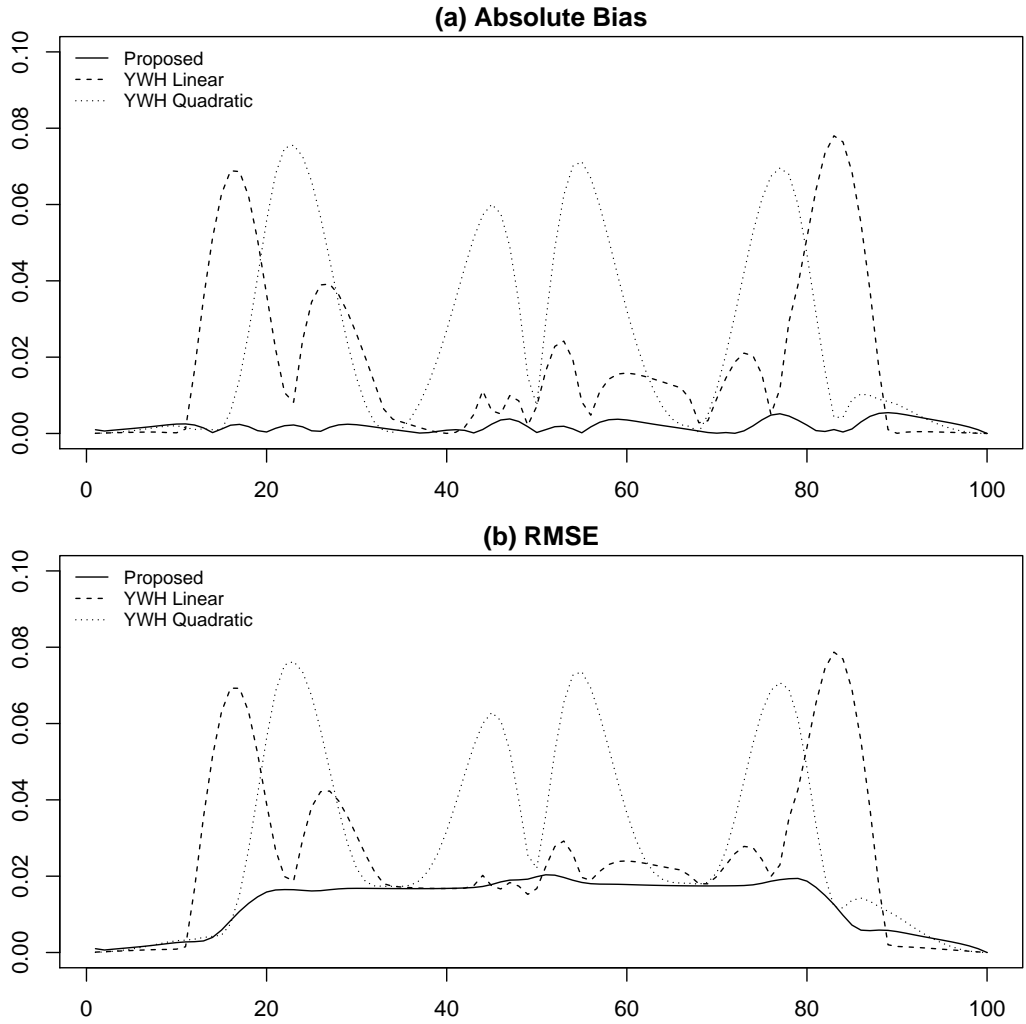}
\caption{Simulation setting 3: (a). absolute bias of $\widehat{\Lambda_0}(t)$. (b). RMSE of $\widehat{\Lambda_0}(t)$.}
\label{fig:sim1}
\end{center}
\end{figure}

\begin{figure}[ht]
\begin{center}
\includegraphics[scale=0.7,angle=270]{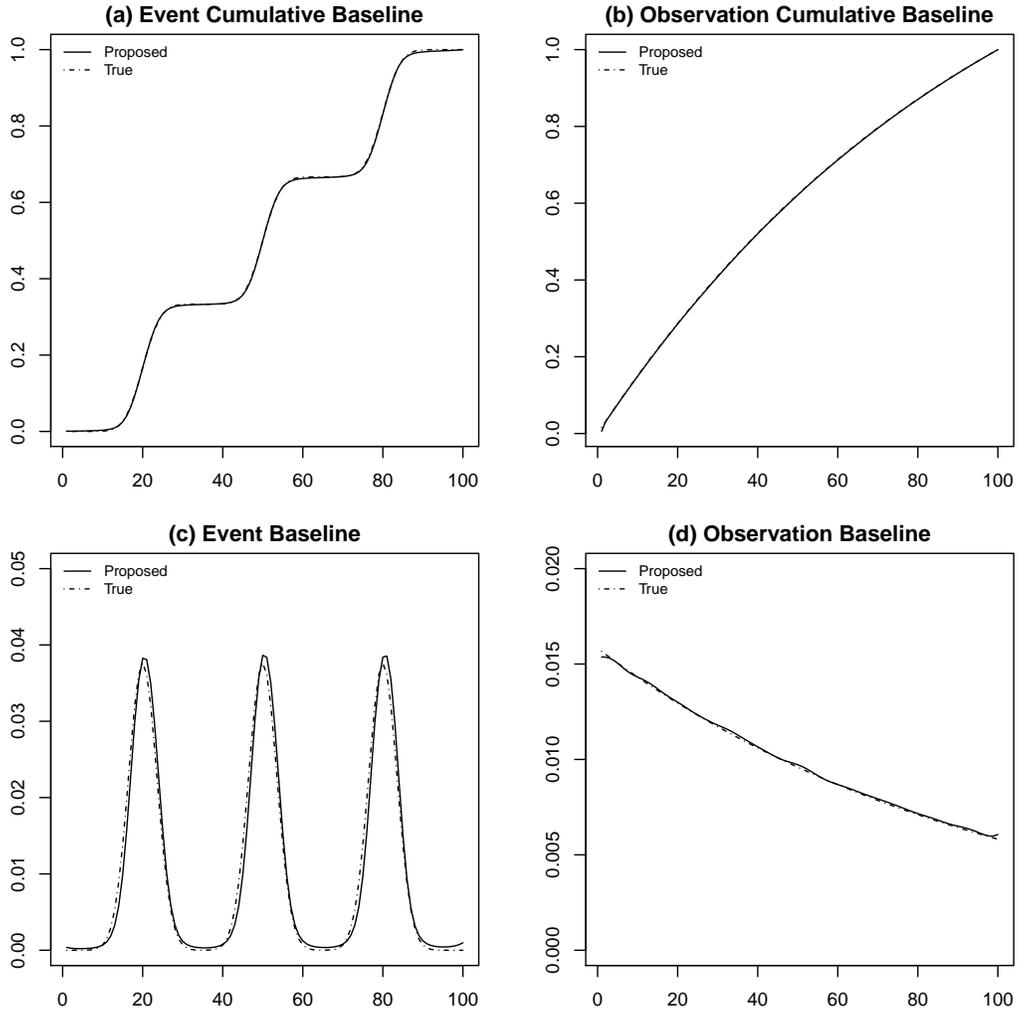}
\caption{Simulation Setting 3: Average fit of 500 replicates. (a). Cumulative function of $\lambda_0(t)$. (b). Cumulative function of $\mu_0(t)$.
	(c). $\lambda_0(t)$. (d). $\mu_0(t)$.}
\label{fig:sim2}
\end{center}
\end{figure}

\begin{figure}[ht]
\begin{center}
\includegraphics[scale=0.8,angle=270]{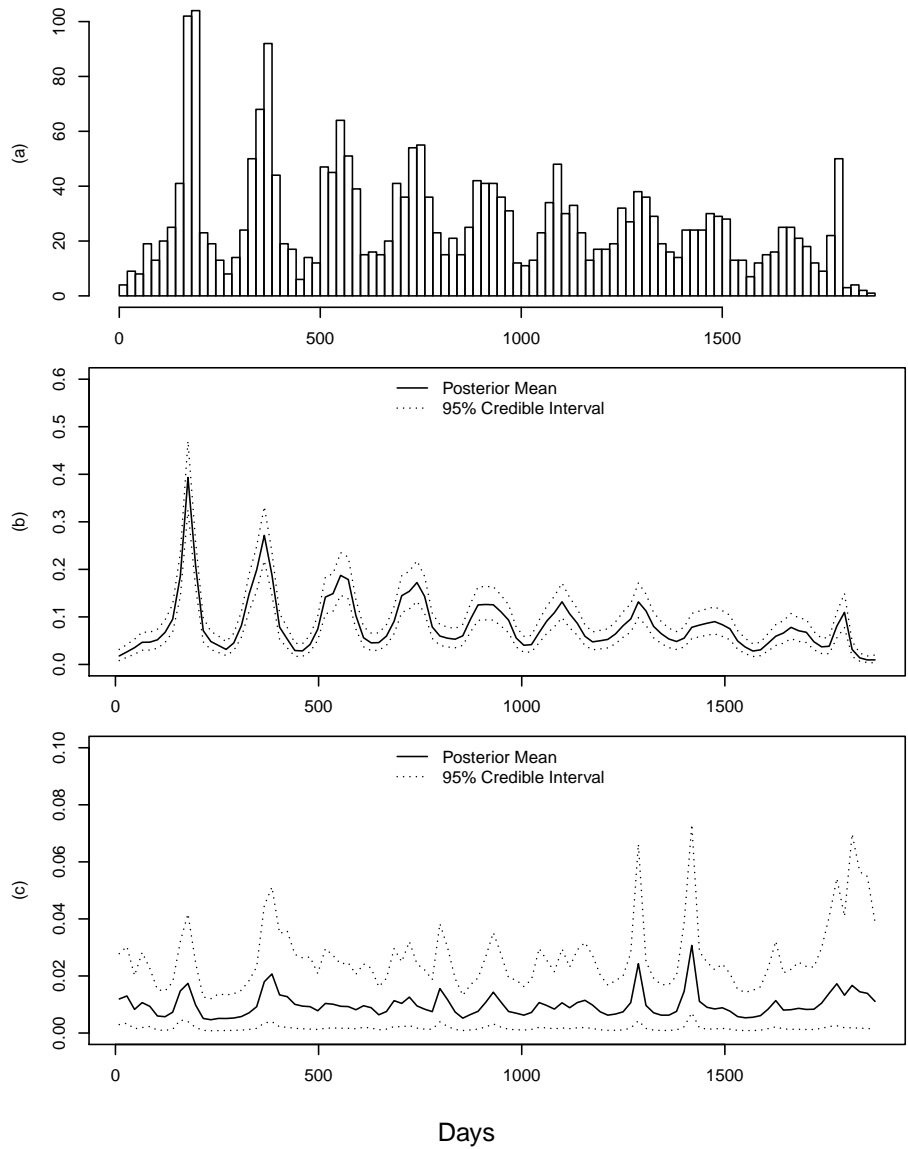}
\caption{Skin cancer data: (a). Actual observation times of all patients. (b). Posterior observation intensity $\mu_0(t)$. (b). Posterior event intensity $\lambda_0(t)$.}
\label{fig:skin1}
\end{center}
\end{figure}

\begin{figure}[ht]
\begin{center}
\includegraphics[scale=0.8, angle=270]{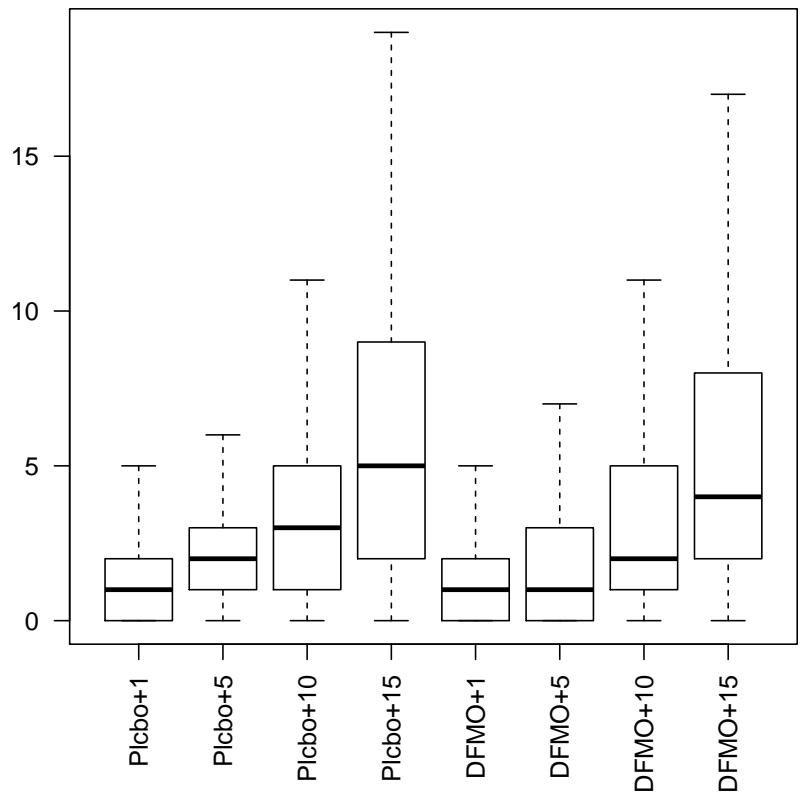}
\caption{Skin cancer data: Predictive distributions for the tumor recurrence count $\tilde{y}$ in 
	five years after treatment. Distributions are displayed for different covariates $\tilde{\bfx}$. 
	``DFMO+5" means the future subject has 5 initial tumors and is treated with DFMO. 
	Outliers beyond $1.5$ Interquartile Quartile Range are not displayed. }
\label{fig:skin2}
\end{center}
\end{figure}

\begin{figure}[ht]
\begin{center}
\includegraphics[scale=0.8, angle=270]{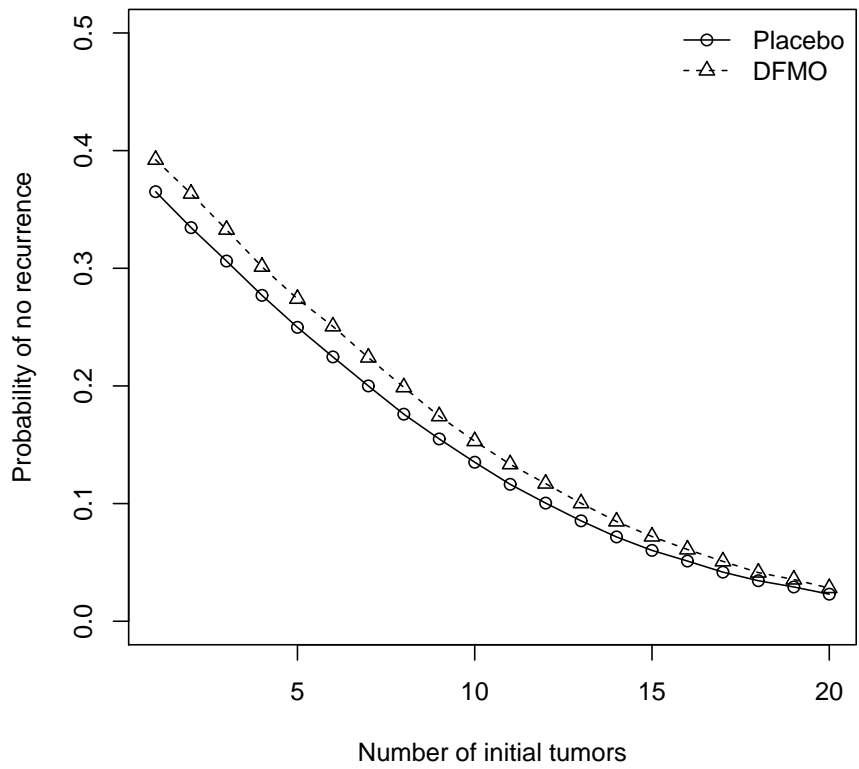}
\caption{Skin cancer data: Predictive probability of no recurrence of skin cancer (disease-free) within five years of treatment:
	$P(\tilde{y}=0 \mid \cfD, \cfT, \tilde{\bfx})$.}
\label{fig:skin3}
\end{center}
\end{figure}

\begin{figure}[ht]
\begin{center}
\includegraphics[scale=0.8, angle=270]{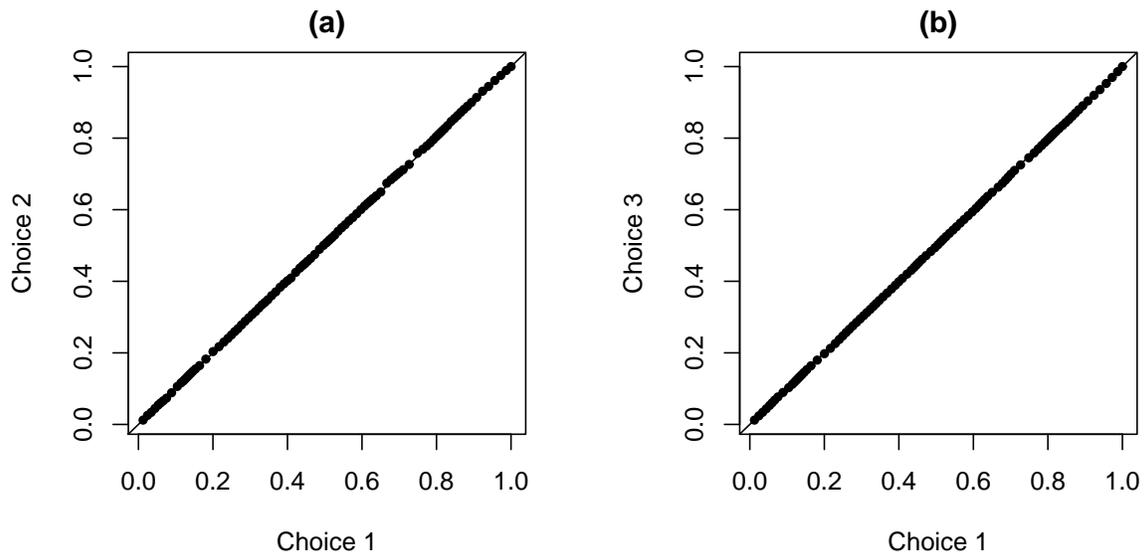}
\caption{Skin cancer data: Sensitivity analysis for the posterior mean estimate of $\Lambda_0(t)$. Compare hyperparameter choices: 
	(1) $\nu=2.5$, $\theta_k\sim\mbox{Ga}(8,4)$; (2) $\nu=1.5$, $\theta_k\sim\mbox{Ga}(4,4)$; and (3) $\nu=0.5$, $\theta_k\sim\mbox{Ga}(16,4)$, 
	$k=1,2$.}
\label{fig:skin4}
\end{center}
\end{figure}